\documentclass[11pt]{article}

\usepackage[latin1]{inputenc}
\usepackage[english]{babel}
\usepackage{amssymb,amsmath}
\usepackage{amsthm}
\usepackage{amsfonts}
\usepackage{geometry}
\usepackage{graphicx}
\usepackage{color}
\usepackage{cite}
\usepackage{multirow}
\usepackage{subfigure}

\theoremstyle{definition}

\newtheorem{thm}{Theorem}

\newtheorem{conj}[thm]{Conjecture}

\newtheorem*{conj*}{Conjecture}
\newtheorem{pbm}[thm]{Problem}

\newcommand{\fobj}{\ensuremath{f}}

\newcommand{\gn}{\mathcal{G}_{n}}
\newcommand{\cn}{\mathcal{C}_{n}}
\newcommand{\kn}{\mathcal{K}_{n}}
\newcommand{\pn}{\mathcal{P}_{n}}
\newcommand{\ks}[2]{K\!S_{{#1},{#2}}}
\newcommand{\hnclass}{\mathcal{H}_{n}}

\newcommand{\maxclass}{\mathcal{M}}
\newcommand{\inv}{\mathcal{I}}
\newcommand{\transf}{\mathcal{T}}
\newcommand{\nor}[1]{\widehat{#1}}

\newcommand{\si}{\Rightarrow}

\newcommand{\ie}{{\emph{i.e., }}}
\newcommand{\eg}{{\emph{e.g., }}}
\newcommand{\etal}{{\emph{et al. }}}

\newcommand{\sloand}[2]{\footnote{N. J. A. Sloan - OEIS Foundation - \texttt{www.oeis.org}, Sequence #1 - #2.}}

\newcommand{\stitle}[1]{\begin{flushleft}\textsc{#1} \end{flushleft}}

\newcommand{\DG}{{\sf Digenes}}

\newcommand{\HOG}{{\sf HoG}}

\makeatletter  

\title{Digenes: genetic algorithms to discover conjectures about directed and undirected graphs}
\author{Romain Absil\footnote{Algorithms Lab, Universit\'e de  Mons, Place du parc 20, B-7000 Mons, Belgium.}\ $^{,}$\footnote{FRIA grant holder.}  \and 
Hadrien M\'elot\footnotemark[1]\ $^{,}$\footnote{Corresponding author. E-mail: {\tt romain.absil@umons.ac.be}.}
}

\begin{document}

\maketitle
\vspace*{0.2cm}

\hrule
\vspace*{0.2cm}
\small
\noindent
\textbf{Abstract.} We present \DG, a new discovery system that aims to help researchers in graph theory. While its main task is to find extremal graphs for a given (function of) invariants, it also provides some basic support in proof conception. This has already been proved to be very useful to find new conjectures since the AutoGraphiX system of Caporossi and Hansen~\cite{AGX1}. However, unlike existing systems, \DG\ can be used both with directed or undirected graphs. In this paper, we present the principles and functionality of \DG , describe the genetic algorithms that have been designed to achieve them, and give some computational results and open questions. This do arise some interesting questions regarding genetic algorithms design particular to this field, such as crossover definition. 

\vspace*{0.2cm}

\vspace*{0.2cm}
\hrule

\normalsize

\section{Introduction} \label{sec_intro}

During the last three decades, several software systems have been dedicated to the task of helping the graph theorist in the process of discovery. This computer aided support is currently much developed to obtain new conjectures or to search for counterexamples. Less efforts (or successes) have been made to assist the writing of proofs and, to the best of our knowledge, there exists no discovery system dedicated to \emph{directed} graphs, all theses systems dealing only with undirected graphs. After settling some notations in Section~\ref{sec_nota}, we briefly survey existing discovery systems in Section~\ref{sec_survey}. 

In this paper, we introduce the new system \DG\  that can deal both with undirected \emph{and} directed graphs. Its principles and functionality are described in Section~\ref{sec_func}. The main goal of \DG\ is to find graphs that are extremal for a given objective function of graph invariants. Moreover, we explain how this system can be used to make conjectures; find graphs satisfying some given constraints; search for counterexamples of a given conjecture; and check graph transformations that can be used in proofs. 

In Section~\ref{sec_ga}, we describe the algorithms and choices we made to develop \DG . These are based on genetic algorithms and do arise interesting questions about graphs encodings and operators (crossovers and mutations) specific to the task of finding extremal graphs.  

We validate our algorithms in Section~\ref{sec_res} with computational results. Finally, we end this paper by drawing concluding remarks and pointing out challenging open questions.

\section{Notations} \label{sec_nota}

This section is devoted to basic definitions and notations used throughout the paper. We assume the reader to be familiar with usual notions of graph theory and we refer to the books of Diestel~\cite{Diestel00} for general graph theory and Bang-Jensen and Gutin~\cite{Bang01} for more details about directed graphs (\emph{digraphs}).

Let $G = (V, A)$ be a simple digraph of order $n(G) = |V|$ and size $m(G)=|A|$. We denote by $\gn$ the space of all simple non isomorphic digraphs of order $n$, and $\nor{\gn}$ the space of all simple non isomorphic undirected graphs of order $n$. A (graph) \emph{invariant} is a numerical value preserved by isomorphism such as chromatic number, independence number, diameter and so on. We also provide some invariants definitions used later.

Recall that the \emph{diameter} $D(G)$ of a graph $G$ is the maximum distance between any pair of its vertices. The \emph{average distance} $\mu(G)$ of a graph $G$ is the arithmetic average of the lengths of all shortest paths of $G$. In the case of digraphs, $G$ must be strongly connected in order to have a finite diameter or average distance. In an undirected graph $G=(V,E)$, the \emph{imbalance} of an edge $\{i,j\}$ is defined as $|d_i - d_j|$, where $d_i$ is the degree of a vertex $i \in V$, and the \emph{irregularity} $irr(G)$ of $G$ as the sum of imbalances of its edges, as firstly stated by Albertson\cite{Albertson97}.

We note $\cn$ and $\kn$ the cycle and the complete graph of order $n$, respectively. We note $\ks{k}{l}$ the undirected complete split graph made by connecting all vertices of a complete graph of order $k$ to all vertices of an empty graph of order $l$.

  Let \fobj\ be an objective function defined by an invariant (or an arithmetic formula depending on some graph invariant -- that is obviously also an invariant), we will consider combinatorial optimisation problems of the form
\begin{align}
\displaystyle \max_{G \in \gn} \fobj(G),\label{eqn:dpbm}\\
\displaystyle \max_{G \in \nor{\gn}} \fobj(G).\label{eqn:upbm}
\end{align}
where $\hnclass \subseteq \gn$ is a class of \emph{digraphs}, and $\nor{\hnclass} \subseteq \nor{\gn}$ is a class of \emph{undirected} graphs. For example, $\hnclass$ can be the class of strongly connected digraphs of order $n$ and $\fobj(G)$ can be defined by $D(G)$.
 
 To simplify presentation, we will only consider maximisation problems since a minimisation problem can easily be translated into a maximisation problem by multiplying the objective function by $-1$.
 
 \section{Existing graph theoretical discovery systems} \label{sec_survey}

Research in mathematics is increasingly aided by computers and graph theory clearly makes no exception since the proof of the 4-color theorem\cite{Appel77, Appel77b, Appel89, Robertson97}.  For example, it is frequent to use graph generators (such as \textsf{geng} of McKay~\cite{Mckay90, Mckay98}) to confront a conjecture to a large set of (small) examples. This enumerative approach can lead to a counter example, or if it is not the case, reinforce the belief that a tested conjecture is true. Also, some conjecture making systems, such GraPHedron~\cite{GphDesc}, use an enumerative step in their underlying processes.

Other systems allow to interactively explore properties of a selected set of graphs, such as invariant's values. It is for instance the case of the pioneering system \textsf{Graph}~\cite{Cvetkovic83d, Cvetkovic81, Cvetkovic94}, its sequel \textsf{newGraph} \cite{newGraph} and \textsf{Grinvin} \cite{Grinvin} (the later one specifically designed for pedagogical purposes). 

The \textsf{Graffiti} system \cite{Fajtlowicz98, Fajtlowicz87, Fajtlowicz88b, Fajtlowicz90, Fajtlowicz95} uses a specific process to maintain a database of counter-examples and conjectures. It has led to hundred conjectures, some of which have attracted much attention. 

Finally, we mention \textsf{AutoGraphiX} (AGX)~\cite{AGX1, AGXsurvey}. The main idea of AGX is to write undirected graph theory problems of type \eqref{eqn:upbm} as combinatorial optimisation problems and then use an heuristic, based on a Variable Neighbourhood search (VNS)~\cite{Hansen01b, Mladenovic97}, to approximate an optimal solution.

In addition to this very short survey, the next section shows other ways that can be used to assist the writing of conjectures and proofs. The interested reader can also find more detailed surveys on conjecture-making systems~in the papers of Hansen \etal \cite{WhatForms, Hansen02}. We observe that all the surveyed systems only deal with \emph{undirected} graphs.

\section{Principles and functionality of \DG} \label{sec_func}

As previously introduced, existing systems offer various features, although they never deal with directed graphs. In this section, we present a new system, called \DG\footnote{Digenes is actually a contraction of the words ``Directed" for directed graphs and ``Genetic"  for genetic algorithms. \DG\ source code is available by request from the corresponding author.}, its basic principles and core features.

We note that while exact methods, such as enumeration used for instance in GraPHedron, might be suited to solve undirected graphs problems of type \eqref{eqn:upbm}, they are most likely inapplicable in the directed case. Indeed, the number of non isomorphic directed graphs\sloand{A000273}{6/12/2012} increases far more quickly than the undirected one\sloand{A000088}{6/12/2012}. For instance, there are $12005168$ undirected graphs of order $10$, while there are $341260431952972580352$ (giving for this order a factor of about $10^{14}$). This growth motivates the use of metaheuristics for these problems, as they allow to find assumed good solutions in reasonable time. As far as we know, VNS is the only metaheuristic that has been ever used to solve such problems, within AGX. Therefore, a natural question is to wonder how another metaheuristic will succeed for this type of problems, and to compare the quality of the approximated solutions or the time elapsed to find them. We give some answers to this question in this paper since \DG\ uses genetic algorithms (GA)~\cite{Talbi01} we specialise to study extremal graph theory. 

A point motivating this choice is that GA are a population metaheuristic. If for some invariant the maximum is actually reached on some class of graphs, a single solution heuristic, such as local search, will only output a single graph while we have to guess the non uniqueness of this solution. On the other hand, population heuristics will output a set of solutions, usually with several assumed extremal solutions within. Also, we note that many algorithmic methods running time grows in particular with optimisation space size. Genetic algorithms do not, their worst case time complexity only growing with population size, linearly\footnote{It actually also grows with the time complexity of its inner components, but so do other metaheuristics.}.

As previously stated by Hansen and Caporossi~\cite{AGX1}, problems modelled as \eqref{eqn:dpbm} and \eqref{eqn:upbm} might seem simple (and restrictive), however, many graph theory problems can be modelled using this formalism. More concretely, \DG\ offers the following features that can be used for discovery, both at the levels of conjectures and proofs :
\begin{enumerate}
\item find extremal graphs for some invariant,
\item find counterexamples to conjectures or graphs satisfying a given set of constraints,
\item check graph transformations.
\end{enumerate}
We detail these features in the following subsections. Note that since \DG\ works with a metaheuristic, all of these features are to be understood as ``assumed optimal". Indeed, a found maximum may well be just local rather than global in the case of extremal graph finding. In the same way, just because no counterexample has been found to a conjecture does not mean it is true. 

As matter of fact, we introduce each of these features by a simple example, and explain then how to generalise it for any similar problem.

\subsection{Extremal graphs finding and constraint handling}\label{subsec:extr-finding}

It is often considered research in extremal graph theory started in 1941 with the result of Tur\`an \cite{Turan}, who wondered what was  simple yet common question in extremal graph theory is to wonder, given a specific invariant \fobj, what is the maximal value that \fobj\ can reach on some graph $G$, possibly subject to constraints. We illustrate by a simple example how \DG\ handles the matter on Problem \ref{pbm:diam}, along with an easy constraint handling.

\begin{pbm}[Directed diameter]\label{pbm:diam}
Let $G$ a strongly connected digraph of order $n$, what are the graphs maximising $D(G)$ ?

\noindent For this problem, one can easily show that 
\begin{equation*}
D(G) \leqslant n - 1,
\end{equation*}
with equality if and only if $G$ has an hamiltonian path with no forward arc.
\end{pbm}

A first naive way to model this problem is to launch the algorithm on the problem ``$\max D(G)$" with no constraints. However, diameter can only be computed on strongly connected graphs\footnote{Actually we are only interested in finite diameter, explaining why we only consider strongly connected graphs}. On the other hand, restricting \DG\ to only work with strongly connected graphs is unwise since non strongly connected graphs might provide good crossover or mutations scheme. Usually, literature recommends to penalise unfeasible individuals in the objective function. Considering these improvements, we can now model Problem \ref{pbm:diam} as follows :

\begin{equation}\label{eqn:diam-model}
\displaystyle \max_{G \in \gn}~\frac{1}{k(G)}\cdot D(G'),
\end{equation}
where $k(G)$ is the number of strongly connected components of $G$, and $G'$ is a graph constructed from $G$ by adding a minimum number of arcs to make it strongly connected.

Of course, there are several ways to add such arcs, with possibly different diameter. This allow us to define a new local optimisation problem in which it might be useful to construct $G'$ maximising diameter.

\begin{figure}[!htpd]
\begin{center}
\begin{tabular}{ccc}
\includegraphics[width=3cm]{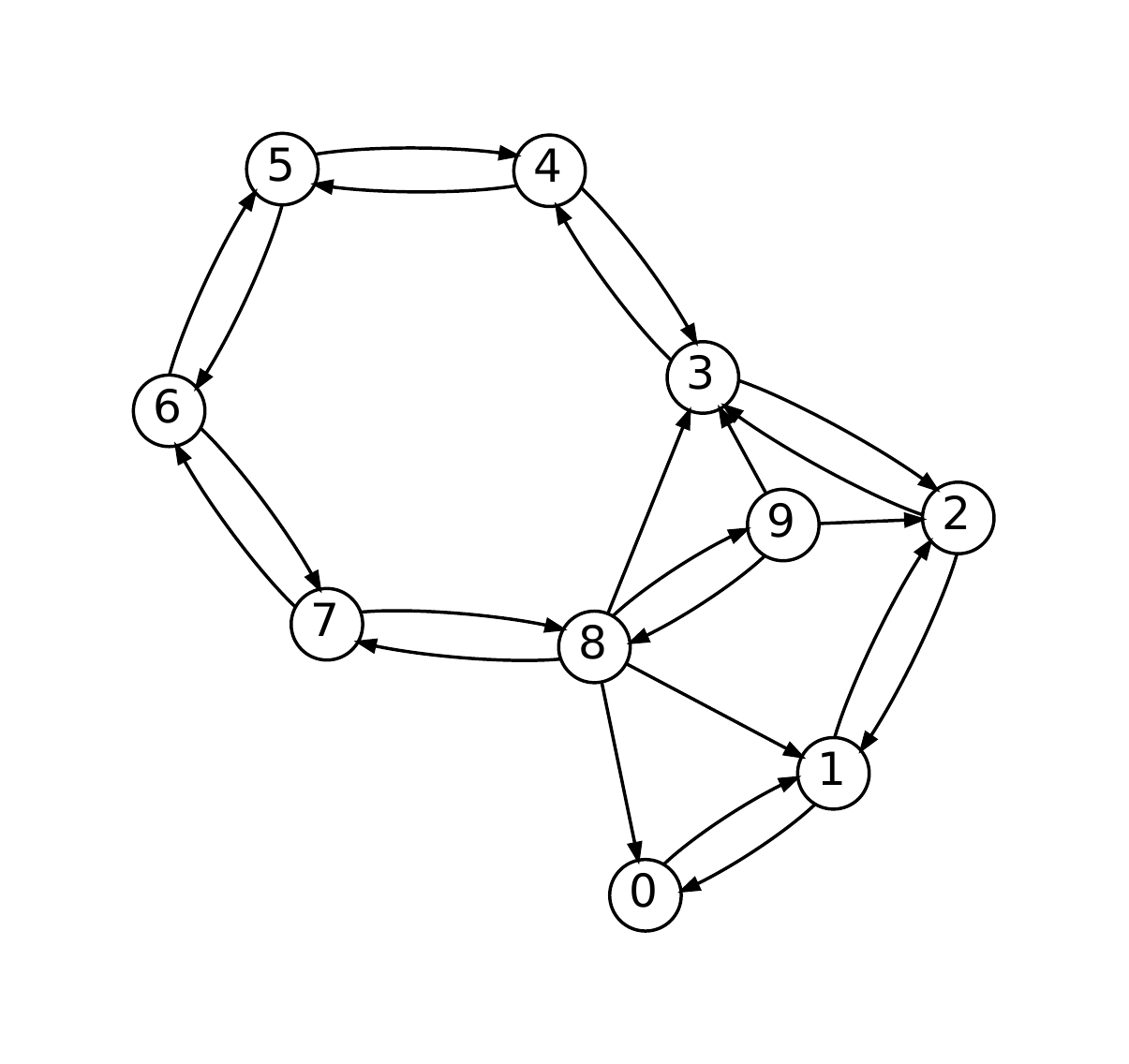} & \includegraphics[width=3cm]{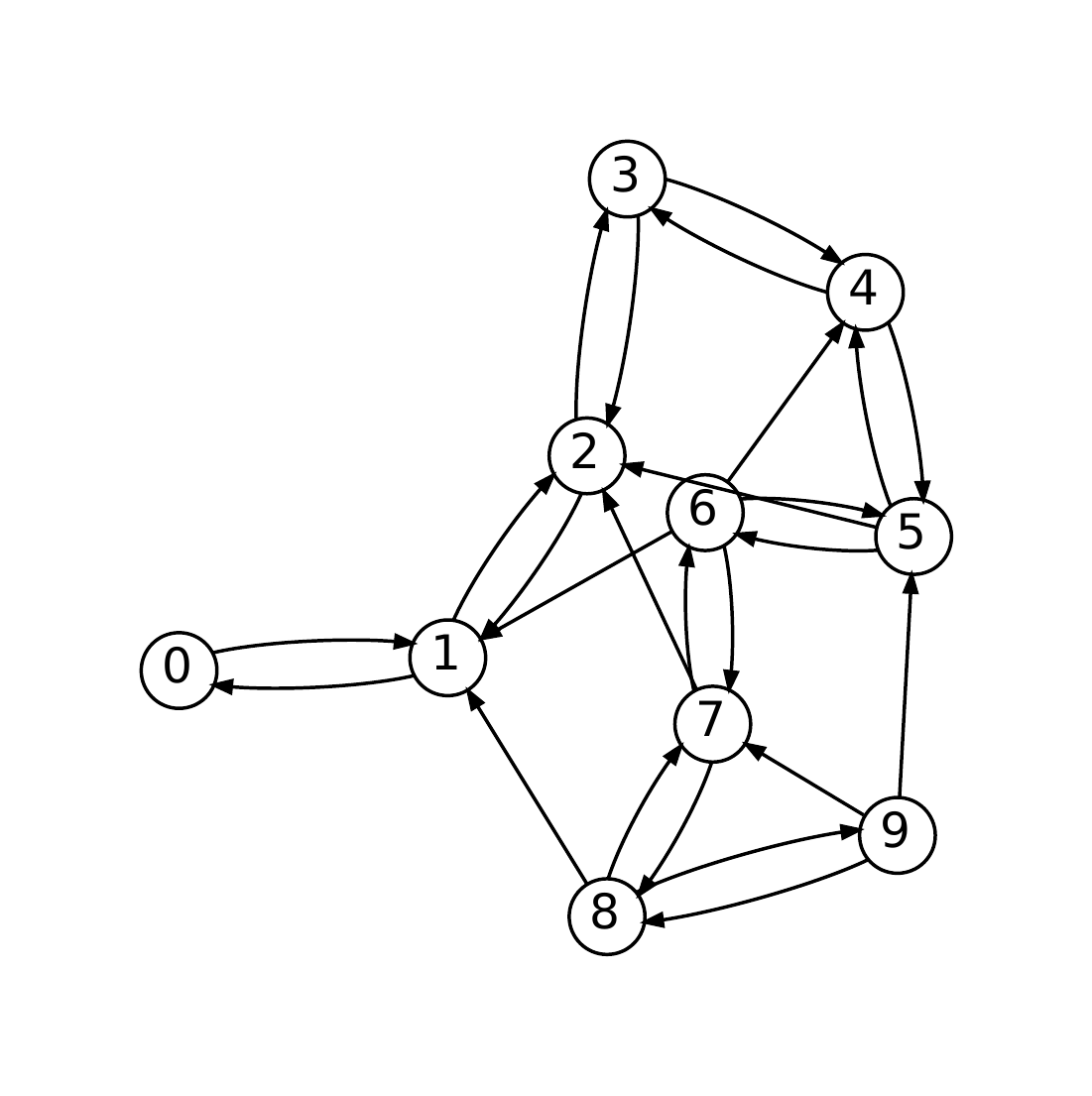} & \includegraphics[width=3cm]{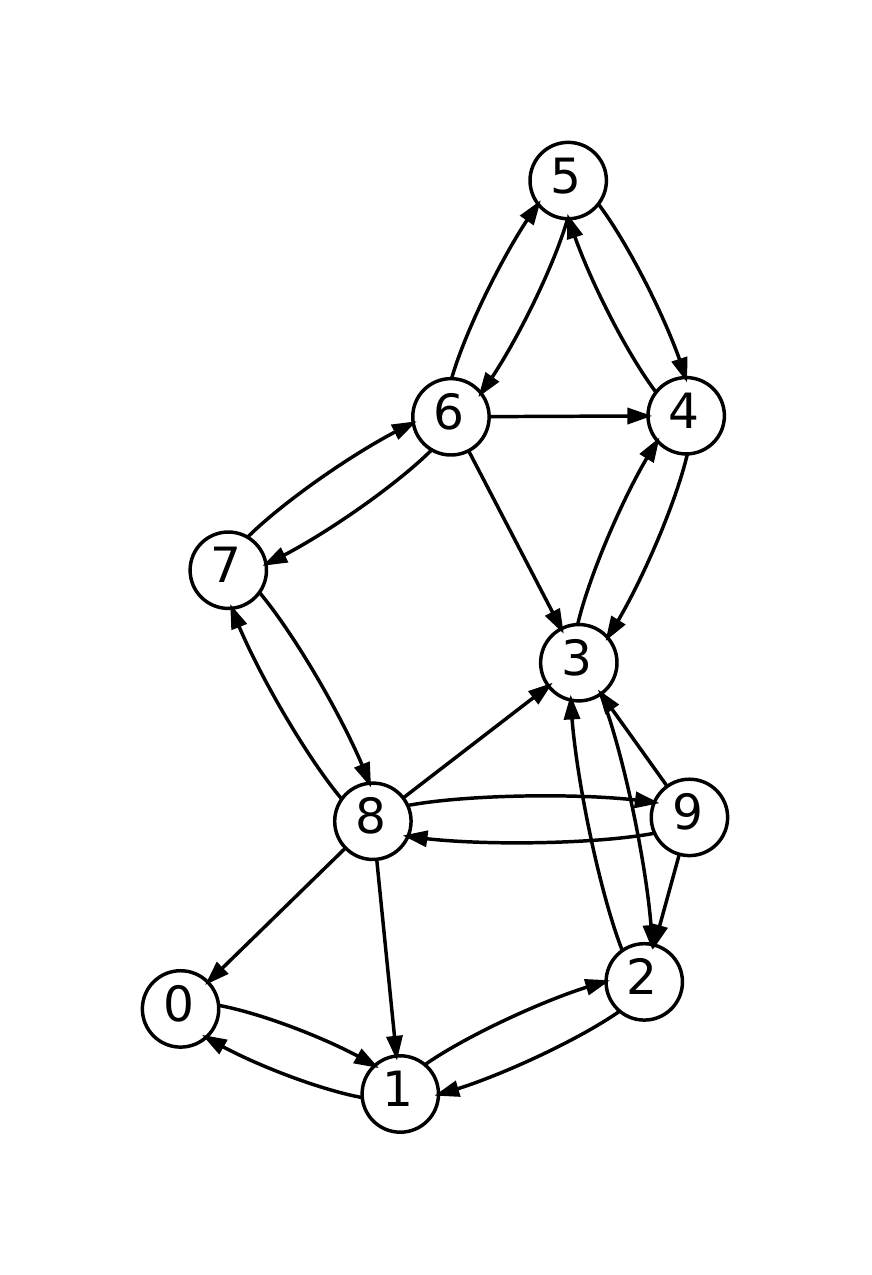} \\
\includegraphics[width=3cm]{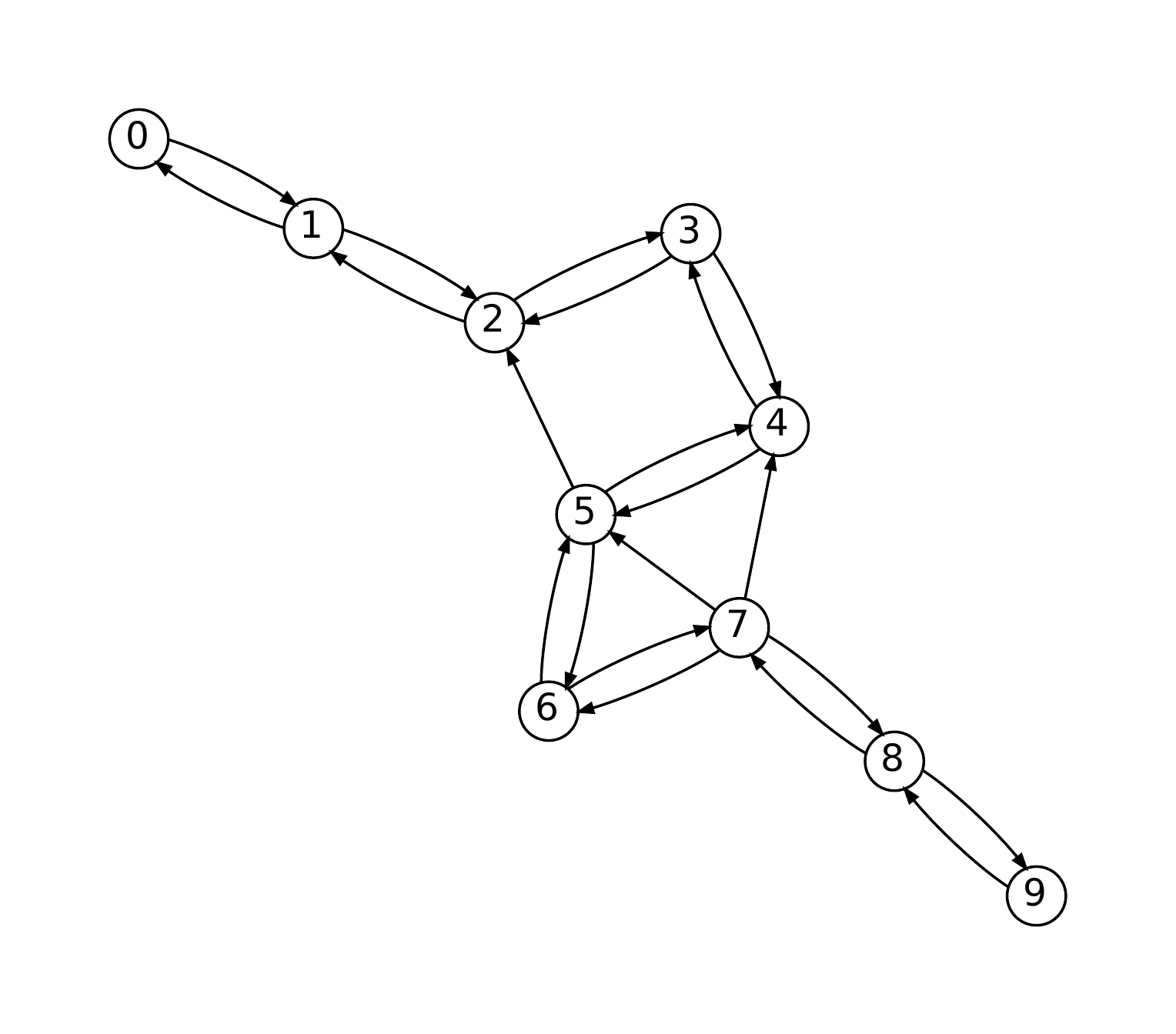} & \includegraphics[width=3cm]{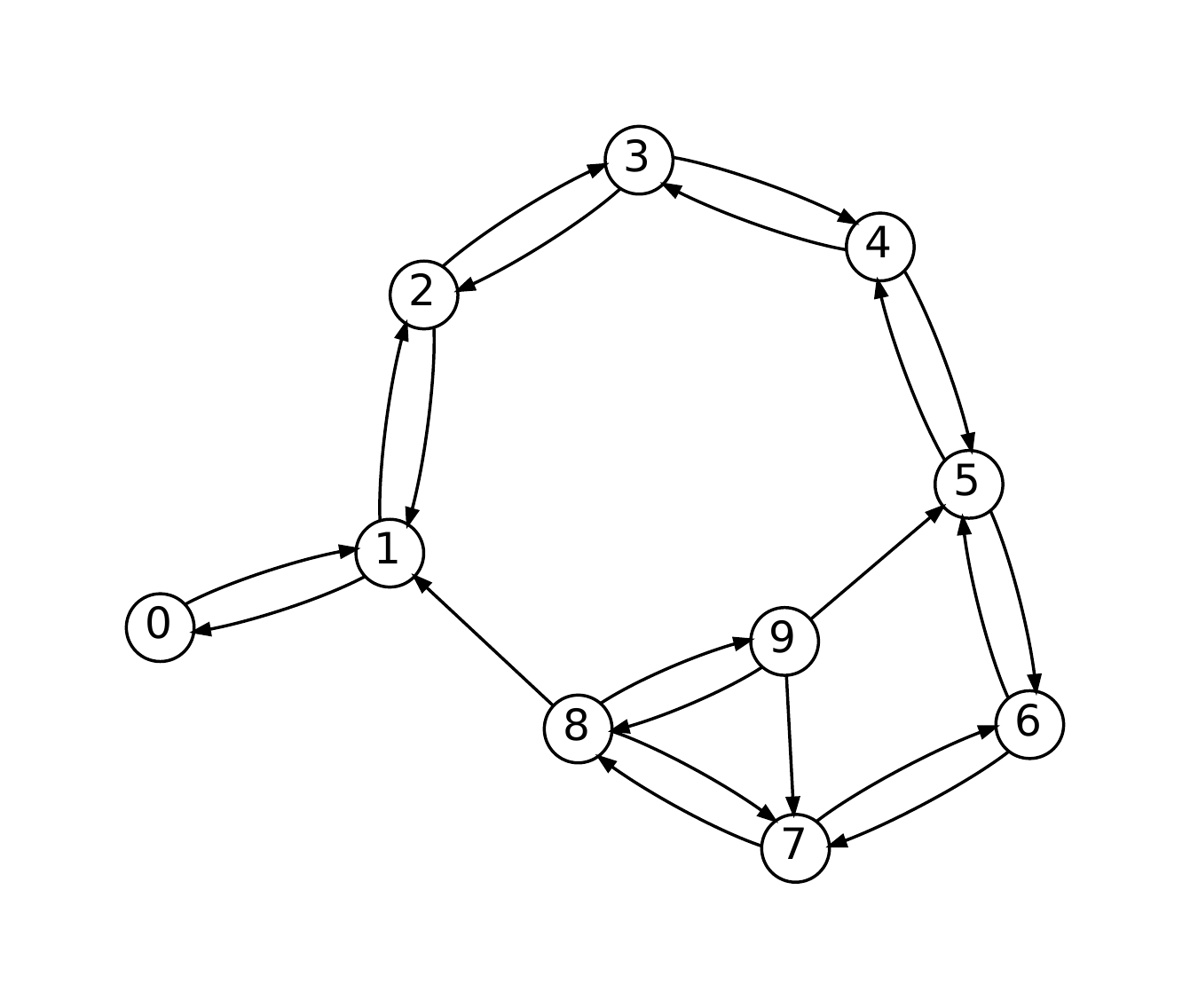} & \includegraphics[width=3cm]{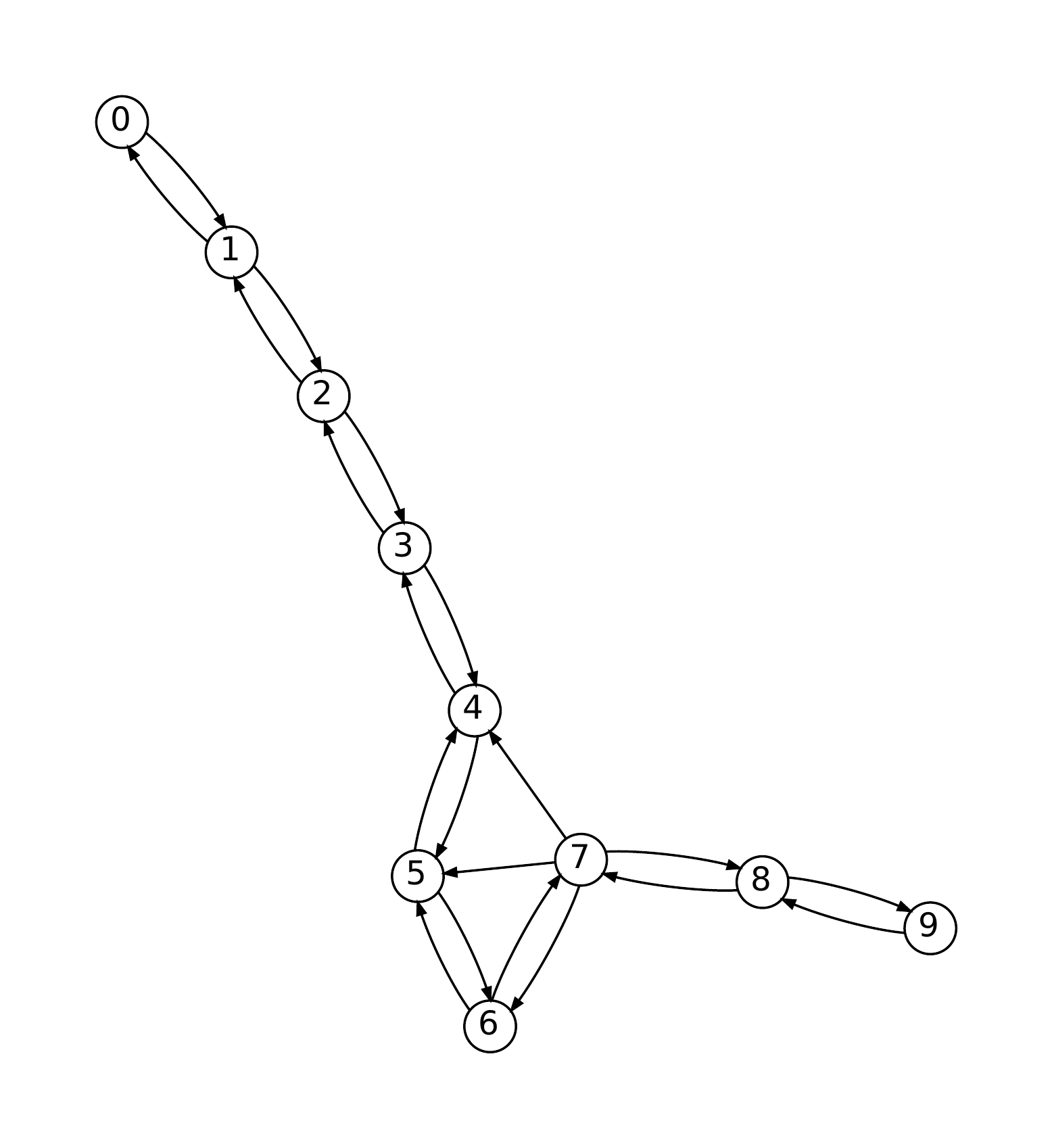}
\end{tabular}
\end{center}
\caption{Sample of output population of Problem \ref{pbm:diam}.}
\label{fig:sample-diam}
\end{figure}

This way, not strongly connected graphs will have a lower objective value, and the "less strongly connected" they are, the less their objective value. When launched on this problem for $n = 10$, \DG\ outputs a population, a sample of which is illustrated in Figure \ref{fig:sample-diam}. We notice these sample graphs are actually extremal graphs described in Problem \ref{pbm:diam}.

We can generalise this approach on any function \fobj\ of invariants. The first step is to handle constraints, then make any graph feasible with a minimum modifications. In our example, we had to ensure the graph was strongly connected by adding a minimum number of arcs (or solve a optimisation subproblem). Finally, assuming \DG\ found graphs belonging to some class $\maxclass$ and maximising \fobj\ reaching a maximum of value $m$, it automatically suggest the following conjecture :

\begin{conj}\label{conj:general}
Let $G \in \gn$, we have
\begin{equation*}
\fobj(G) \leqslant m
\end{equation*}
with equality if and only if $G \in \maxclass$.
\end{conj}

Of course the definition of $\maxclass$ will depend on some generalisation of what \DG\ actually found, for instance if all extremal found graphs are trees, we will assume $\maxclass$ is actually the set of all trees and not only the set the found ones. On the other hand, if there is a unique extremal graph instead of a class, we will assume equality holds if and only if $G$ is isomorphic to this graph rather than belonging to some class. For instance, it is well known that diameter is maximum on undirected graphs only on the path $\pn$. For this problem, \DG\ actually outputs a population consisting f many isomorphic copies of $\pn$. We will see in Section \ref{sec_res} examples where only one graph is extremal.

\subsection{Finding graph satisfying constraints or counterexamples}

Given a set of constraints of type $\inv(G) \leqslant \inv'(G)$ or $\inv(G) = \inv'(G)$, where $\inv$ and $\inv'$ are both graph invariants, one can simply wonder if there exists graphs satisfying these conditions. Moreover, if there are, how can we characterise these graphs ? It is relevant consider such constraints since Hansen \etal \cite{WhatForms} stated that most of conjectures and results in graph theory are equalities or inequalities between graph invariants.

The following simple example illustrates how we can solve such types of problems.

\begin{pbm}\label{pbm:undir-conn}
Is there any undirected graph of order $n$ and size $m$ such as $m = n - 1$ ?

\noindent It is trivial there are, however we are more interested in characterise these graphs. In that case, one can easily show that all undirected trees, and only them, match the property.
\end{pbm}

We can simply model this problem in \DG\ by ``$\max -|m - n + 1|$". If a graph have more or less edges than vertices, it will have a negative objective value. Otherwise, if the constraint is respected, it has a zero objective value.

We can generalise the handling of multiple constraints of type $\inv_k(G) = \inv'_k(G)$ simply by adding the terms $-|\inv_k(G) - \inv'_k(G)|$ to the objective function. We handle inequality constraints in a similar way.

This simple tool can also be used in order to find counterexamples to check, reinforce or reject conjectures. For instance, if we have a conjecture of the same type as Conjecture \ref{conj:general}, we simply have to maximise $\fobj(G) - m$. If a graph $G$ is found with $\fobj(G) - m > 0$, the conjecture is rejected.

\subsection{Help for proofs and transformation validation}

A common problem in extremal graph theory is to prove some graph $G^*$ to be optimal for some invariant $\inv$, \ie $\forall G \in \gn, \inv(G) \leqslant \inv(G^*)$. A simple yet powerful approach to solve this problem is to use transformation proof. Intuitively, the idea is to find a finite sequence of $k$ graph transformations $\transf_i$ that will eventually end up on $G^*$ by always increasing the value of $\inv$. More formally, at each step and for any graph $G$ and some graph transformation $\transf_i$ (for $1 \leqslant i \leqslant k)$, we want to ensure that
\begin{equation}\label{eqn:valid}
\inv(G) < \inv(T_i(G)).
\end{equation}

For instance, Hansen \etal \cite{AvgDistForest} used $5$ graph transformations increasing $\mu(G)$ while keeping $F(G)$ unchanged, where $F(G)$ is the order of a maximal induced forest.

A key question in transformation proof is then, given a graph transformation $\transf_i$, to wonder if Property \ref{eqn:valid} holds. If it does, we say that $\transf_i$ is \emph{valid}. \DG\ provides help in this sense.

Indeed, while the system will not suggest a valid transformation (that is, a working one), given a transformation, \DG\ will prove it invalid by automatically finding a counterexample, or will conjecture it to be valid when unable to find such a counterexample.

Again, just like in the previous sections, simple use of combinatorial modelling with the problem ``$\max \inv(G) - \inv(\transf(G))$" solves the problem. At the end of the optimisation process, if some $G$ is found among the population with a positive value of objective function, then it is a counterexample and the transformation is not valid. Moreover, counterexamples to an invalid transformation are still useful since they can serve as basis for the construction of another transformation.

On the other hand, a common practice in graph transformation is to wonder whether some properties are kept under graph transformation. More concretely, given a property $P$ and a transformation $\transf$, one can simply ask the following question : if $P$ holds on some graph $G$, does it always holds on $\transf(G)$ ? This is a second main feature in \DG\ regarding transformation validation. 

The system proceeds basically in the same way, by modelling the problem under a combinatorial optimisation problem. Then the system is able to reject the transformation if a counterexample is found during the optimisation process, or conjecture it valid otherwise. This feature is illustrated in the following example.

\begin{pbm}\label{pbm:valid}
Let $G \in \gn$, and $P(G)$ the following property : $\inv_1(G) \leqslant \inv_2(G)$, where $\inv_1$ and $\inv_2$ are two graph invariants. Given some graph transformation $\transf$, we want to check whether
\begin{equation*}
\inv_1(G) \leqslant \inv_2(G) \si \inv_1(\transf(G)) \leqslant \inv_2(\transf(G))
\end{equation*} 
holds on any graph $G$. This question can be simply modelled by the following problem :
\begin{equation*}
\begin {array}{l}
\max \fobj(G)\\
\textrm{where }\fobj(G) = \left\{
   \begin{array}{ll}
   \displaystyle \max \inv_1(\transf(G)) - \inv_2(\transf(G)) & \textrm{ if } \inv_1(G) \leqslant \inv_2(G),\\[0.5em]
   \displaystyle \max \frac{\inv_1(\transf(G)) - \inv_2(\transf(G))}{\inv_1(G) - \inv_2(G)} & \textrm{ otherwise.}\\
   \end{array}
\right.\\
\end{array}
\end{equation*}
As before, graphs with positive value of objective function are counterexamples, and penalising non feasible graphs in the objective function improve genetic algorithm effectiveness.
\end{pbm}

In practice, of course, the objective function design widely depends on the property you want to know if it's kept under graph transformation.  Moreover, various types of properties can be modelled in the way it is in Problem \ref{pbm:valid}, thus allowing to study these type of questions.

\section{A genetic algorithm for extremal graphs} \label{sec_ga}

In this section, we describe exactly what genetic algorithm components are implemented in \DG, as well as how they are. We hereby assume the reader familiar with general genetic algorithm concepts\cite{Talbi01}. We start with several initial population generators, follow with crossover and selection operators, and finish with mutations.

\stitle{Initial population generators}

A first and core component of genetic algorithms is the initial population from which you start, an assumed ``well spread" sample of the studied optimisation space. There are several standard generations routines\cite{Talbi01}[pp. 193-198], we present here four of those available in \DG, namely
\begin{itemize}
\item random generator,
\item random degree sequence generator,
\item sized block generator,
\item House of Graphs (HoG) \cite{Brink03}.
\end{itemize}

A first idea in order to generate random graphs is to generate random encodings of graphs. Random generation and random degree sequence generation work in that way. The first one randomly generates binary adjacency matrices, the second one random degree sequences, and then map that coding to a given graph. In random generation, each arc has then a given probability to be in the graph, in random degree sequence generation, each vertex has a randomly generated in and out-degree. We use the Havel-Hakimi algorithm\cite{Havel01,Hakimi01} to map degree sequences to directed graphs. If the mapping algorithm fails, the degree sequence is dropped and a new one is generated.

In sized block generation, the basic idea is to spread graph size. That is, at order $n$, for each possible size $m \in [0, n(n-1)]$, we generate $k_m$ graphs. In order to respect graph size distribution as much as possible and in polynomial time, we use their known repartition for small orders\sloand{A052283}{06/12/2012}, and extrapolate this repartition for bigger graphs. Then, we only have to generate graphs in blocks of fixed size $k_m$.

Finally, HoG \cite{Brink03} is a small database of "interesting" graphs, for instance complete graphs, cycles, stars, snarks, trees, etc. The idea is then to explicitly add some of these graphs in the initial population, along with some randomly generated graphs (with one of the previous schemes). We will see in Section \ref{sec_res} that this simple feature might vastly increase the algorithm performances.

\stitle{Selection operators}

Selection methods are a core concept in genetic algorithm, since they select which individuals mate and who survives in the next generation. \DG\ implements several of standard operators existing in literature \cite[pp. 205-221]{Talbi01}. 

More particularly, \DG\ offers four main selection methods : roulette wheel selection, stochastic universal sampling, tournament selection and direct elimination. 

Intuitively, in roulette wheel strategy, each individual has a probability directly proportional to its fitness to be selected. Stochastic universal sampling proceeds in a similar way, with lower bias. In tournament selection, a number $k$ of individuals is randomly chosen from the population, the best of these ones is output. This procedure is repeated as many times as needed. We note that adjusting wisely the value of the parameter $k$ usually have a significant impact on the algorithm convergence \cite{Goldberg01}. Finally, in direct elimination, individuals are placed randomly on a direct elimination grid, and individuals are selected after computing the winners.

\stitle{Crossover schemes}

Crossover operators define most of the evolution behaviour of a genetic algorithm, \ie how to mate parents and output siblings. \DG\ allows the use of various types of standard crossovers operators, like \textsf{k-Point}s crossover, as well as some ones more particular to graphs, like {\sf Align-Greedy}. 

In k-points crossover, parent graphs are encoded under integer array format, then $k$ cross-points are randomly chosen. Offsprings are computed by alternatively picking parts of the two parents encodings. For this purpose, two types of encodings are provided, although they basically work the same way. The first one simply put the adjacency matrix of a directed graph into a one dimensional array. The second one does the exact same thing, but sorts vertices by decreasing in and out-degrees. For undirected graphs, only the upper triangular part of the adjacency matrix is encoded.

While this idea is simple, we note a slight problem regarding graph encoding : two isomorphic graphs can have different encodings, that is, the encoding operator is sensitive to vertex labelling. More concretely, if two isomorphic graphs mate, they might give birth to two non isomorphic children. Since isomorphism problem solving is difficult \cite{Garey79}, a basic idea is to use heuristics to solve this disparity while avoiding long computations.

A simple yet powerful approach is simply to use local search on the graph adjacency matrix. More formally, given two matrices $A$ and $B$, the algorithm swaps rows and columns $i$ and $j$ in $B$ if it increases the number of matching entries between $A$ and $B$, and repeat that process as many times as possible. When such a permutation cannot be found, we said that the two corresponding graphs are \emph{aligned}.

A first simple crossover scheme is then to apply k-points crossover on two aligned graphs. We note that crossover \textsf{Align-KPoints}. Another idea is, with aligned graph $G_1$ and $G_2$, to alternatively pick good subgraphs from $G_1$ and $G_2$ to build children. Of course, the way subgraphs are picked depends on the studied problem. For instance, we can use a greedy heuristic with the underlying fitness to choose ``the best" way to pick a vertex from one of these graphs. We call this crossover scheme \textsf{Align-Greedy}



\stitle{Mutations}

Basically, a mutation is a function modifying a graph in order to promote diversity, for instance by adding, removing or moving some of its edges. Each generation, some individuals mutate (with an assumed low probability) to produce new, different individuals.

\begin{figure}[!htpd]
\begin{center}
\begin{tabular}{cccc}
\subfigure[Add edge]{\includegraphics[height=1cm]{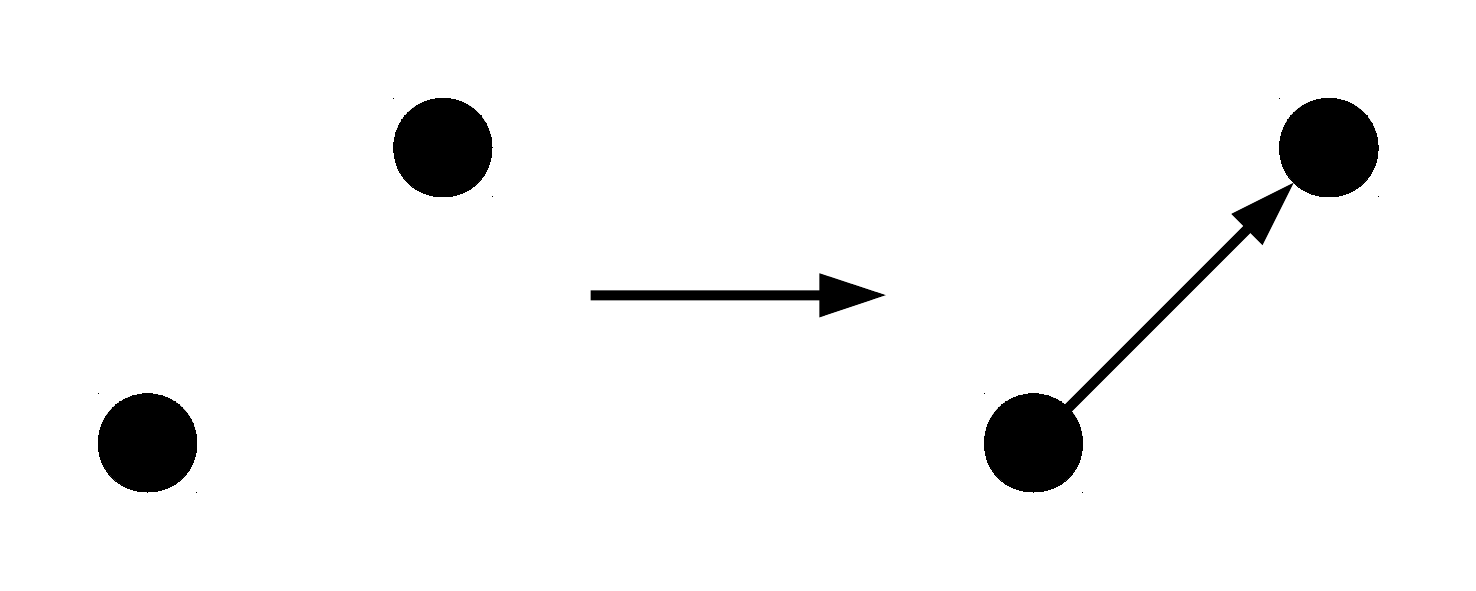}} & \subfigure[Reverse edge]{\includegraphics[height=1cm]{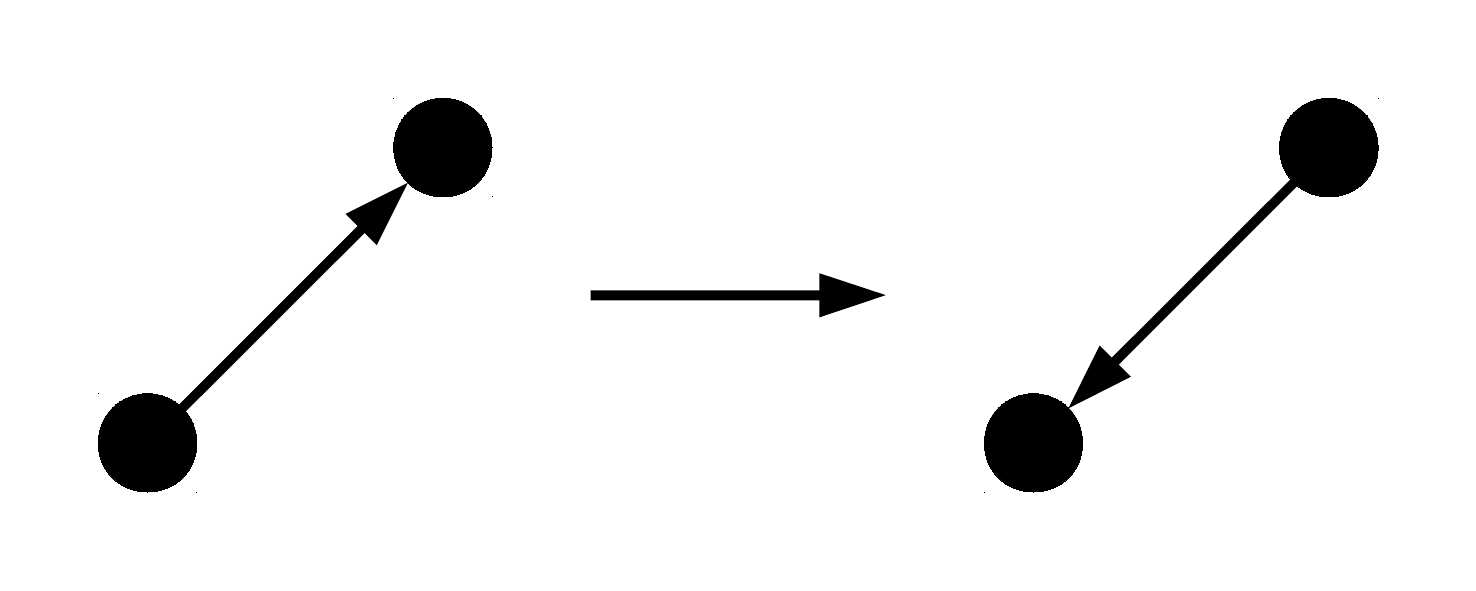}} & \subfigure[Shortcut]{\includegraphics[height=1cm]{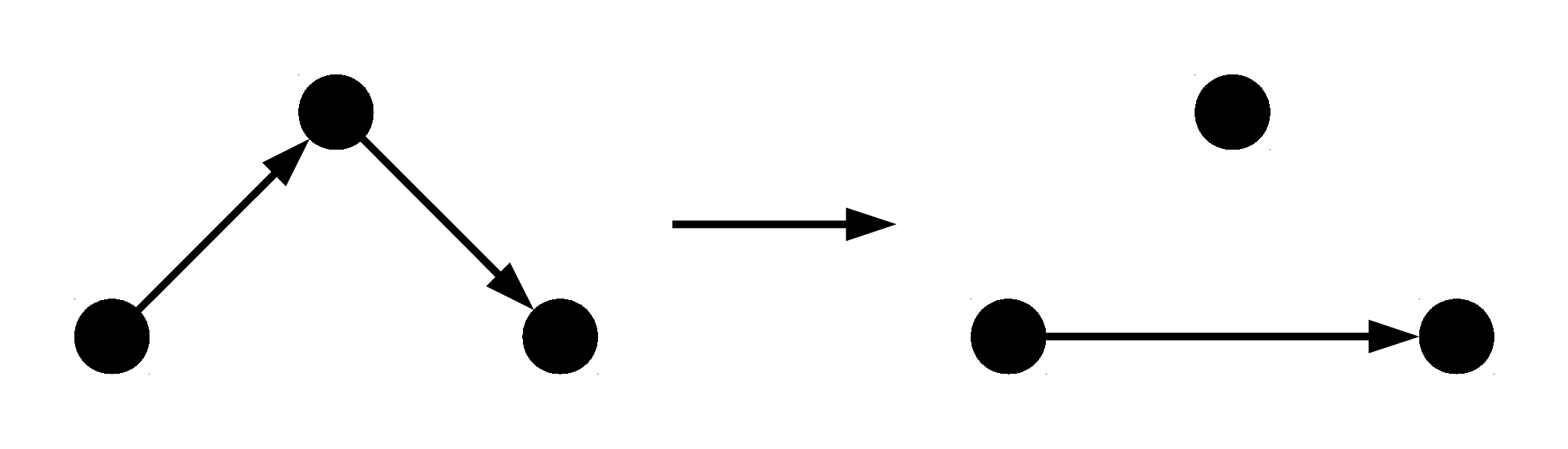}} & \subfigure[Move edge]{\includegraphics[height=1cm]{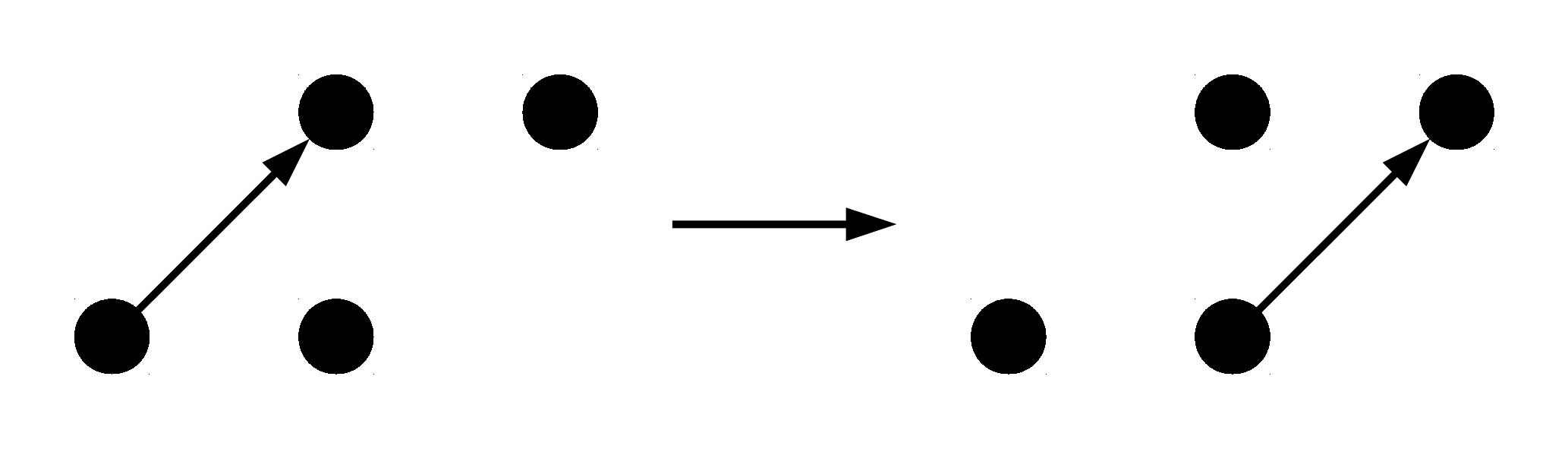}}\\
\subfigure[Remove edge]{\includegraphics[height=1cm]{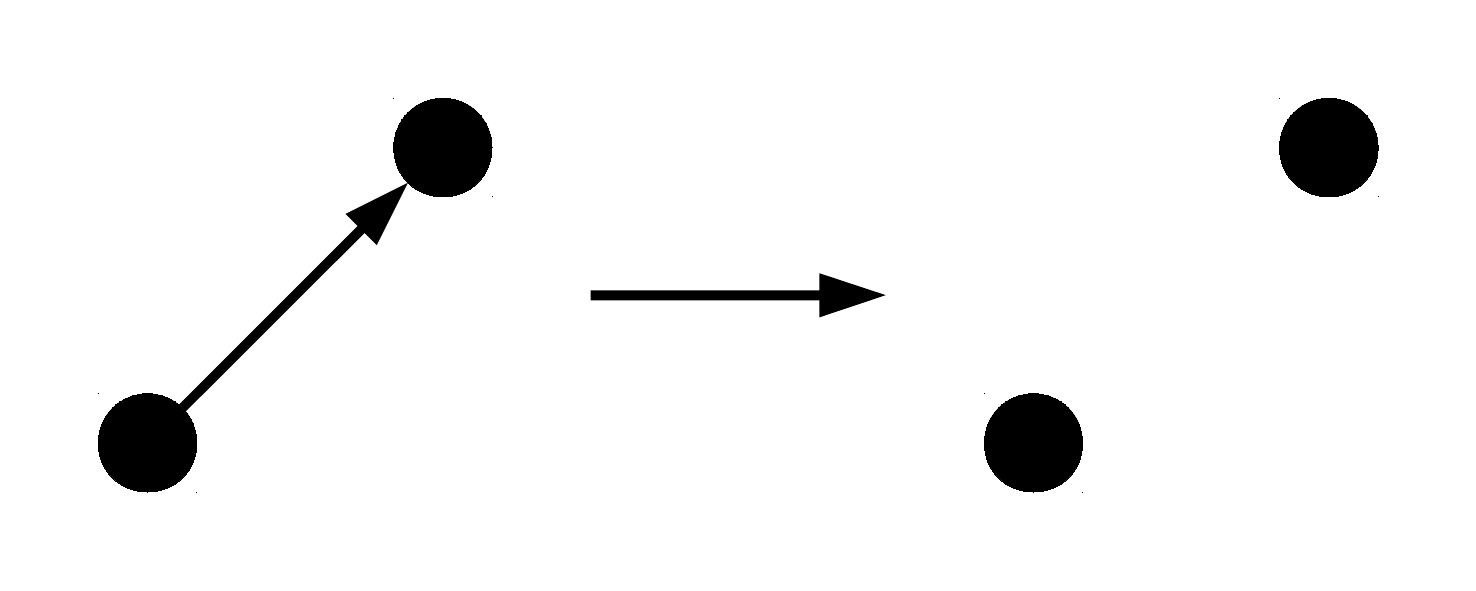}} & \subfigure[Rotate edge]{\includegraphics[height=1cm]{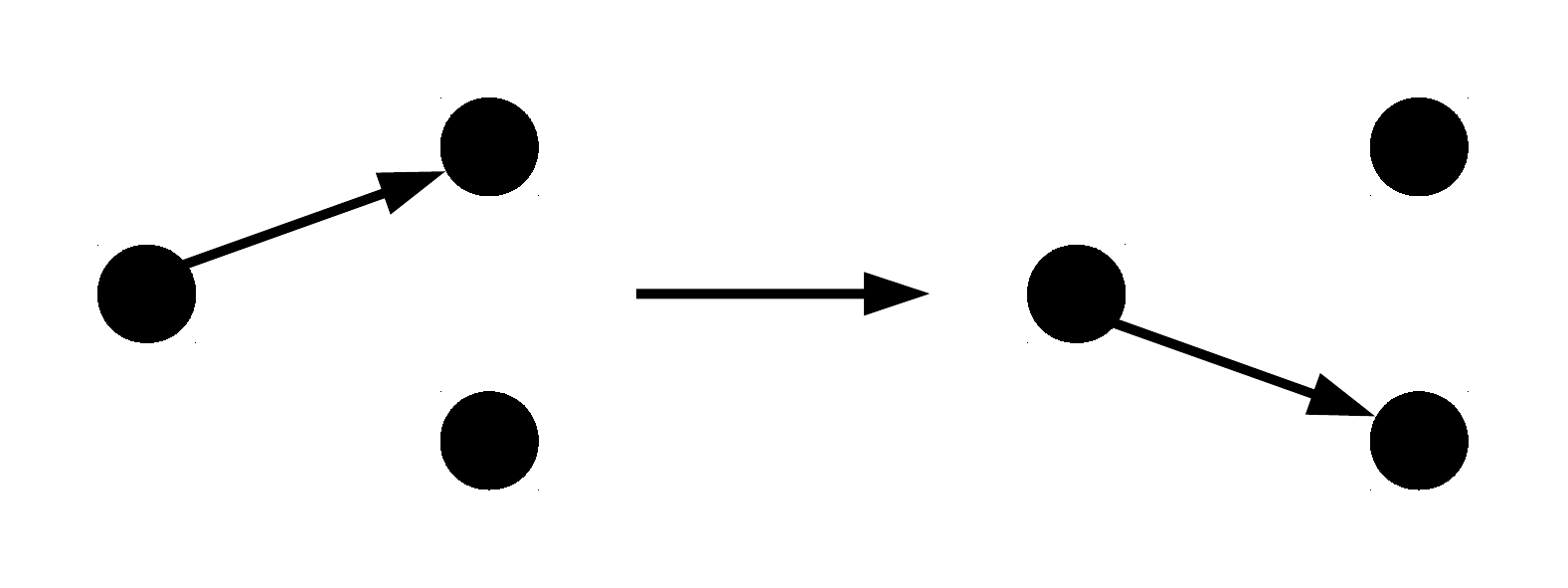}} & \subfigure[Detour]{\includegraphics[height=1cm]{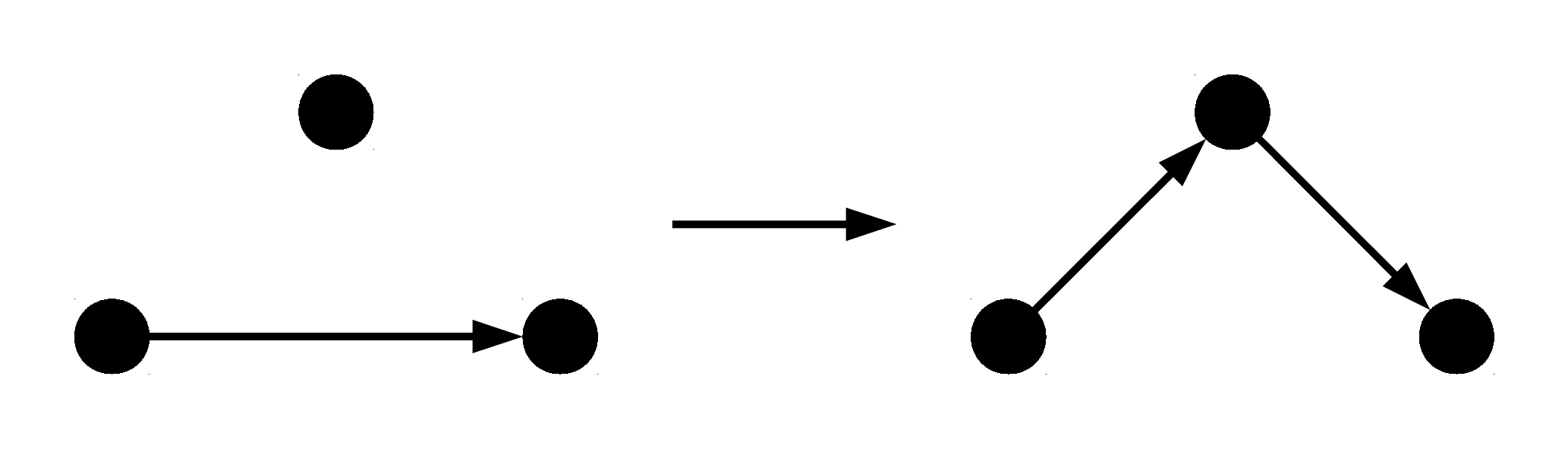}} & \subfigure[2-Opt]{\includegraphics[height=1cm]{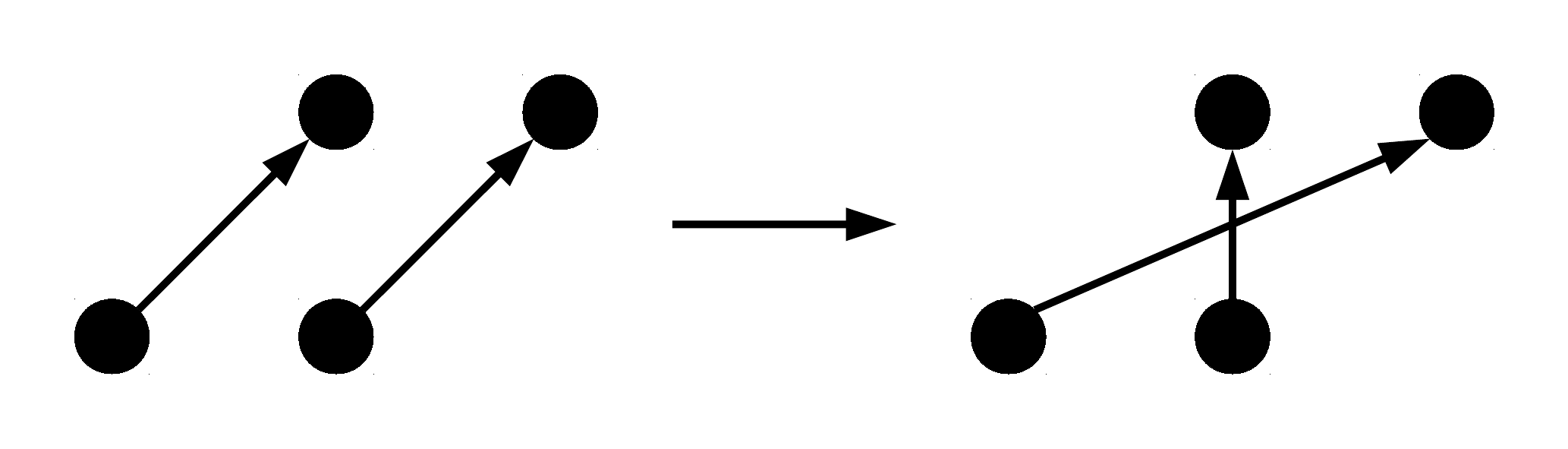}}\\
\end{tabular}
\end{center}
\caption{Illustration of graph mutations}
\label{fig:muta}
\end{figure}

\DG\ offers various general (small) graph mutation schemes, inspired from local search moves we can for instance find in AGX\cite{AGX1} and that we particularise for directed graphs. Figure \ref{fig:muta} illustrates some of these mutations. In order to reinforce the mutation factor, it is sometimes useful to chain some of these schemes multiple times in the same mutation process.


\section{Computational results} \label{sec_res}

In the following section, we present some computational results in order to validate \DG\ and its underlying algorithms. For this purpose, we consider Problem \ref{pbm:diam}, along with the two following ones. 
%


\begin{pbm}[Directed average distance]\label{pbm:avg}
Let $G$ a strongly connected digraph of order $n$, what are the graphs maximising $\mu(G)$ ?

\noindent For this problem, Doyle and Graver~\cite{doyle02} proved that 
\begin{equation*}
\mu(G) \leqslant \frac{n}{2},
\end{equation*}
with equality if and only if $G \simeq \cn$.
\end{pbm}

\begin{pbm}[Undirected irregularity]\label{pbm:irreg}
Let $G$ an undirected graph of order $n$, what are the graphs maximising $irreg(G)$ ?

\noindent Albertson~\cite{Albertson97} showed that 
\begin{equation*}
irreg(G) \leqslant \frac{4n^3}{27},
\end{equation*}
although this bound is not tight. Moreover, Hansen and M\'elot~\cite{agx9} proved that irregularity is maximal if and only if
\begin{equation*}
G \simeq \left\{
\begin{array}{ll}
\displaystyle K\!S_{\frac{n-2}{2}~,~\frac{n+2}{2}} & \textrm{ if } n \textrm{ is even,}\\[0.5em]
\displaystyle K\!S_{\frac{n-3}{2}~,~\frac{n+3}{2}} & \textrm{ otherwise}.
\end{array}\right.
\end{equation*}
\end{pbm}

Table \ref{tab:stats} denotes statistics regarding these three problems. Each problem has been launched a hundred times, during 20000 generations, with stochastic universal sampling selection both for parents and survivors selection, \textsf{Align-2Point} crossover and an initial population generated in blocks of fixed size along with some graphs picked from \textsf{HoG}.

For each of the subtables, first column denotes the order of computed graphs, the second one being the \emph{raw success} (R. S.) of the algorithm, that is, the number of times the algorithm actually found the optimum, while the third one denotes the \emph{amortised success} (A. S.), \ie the average of ratios between the maximum found fitness and the optimum fitness. The fourth column gives the average time (t.(s)), in CPU seconds\footnote{Output from an Intel \textcopyright\ Core \texttrademark\ i5 420M @ 2.53GHz.}, to run one test, and the last one the average generation number (t.(gen)) on which the optimum is firstly found, when it is found.

\begin{table}[!htpd]
\begin{center}
\scriptsize{
\begin{tabular}{|c||r|r|r|r||r|r|r|r||r|r|r|r|}
\hline
\multirow{2}{*}{$n$}& \multicolumn{4}{c||}{\textbf{Directed diameter}} & \multicolumn{4}{c||}{\textbf{Directed average distance}} & \multicolumn{4}{c|}{\textbf{Undirected irregularity}} \\
\cline{2-13}
 & R. S. & A. S. & t.(s) & t.(gen) & R. S. & A. S. & t.(s) & t.(gen) & R. S. & A. S. & t.(s) & t.(gen) \\
\hline
5 & 100 & 1.0 & 15.3 & 17.8 & 100 & 1.0 & 16.5 & 48.3 & 100 & 1.0 & 6.3 & 16.3 \\
10 & 100 & 1.0 & 26.3 & 85.4 & 100 & 1.0 & 34.6 & 99.1 & 100 & 1.0 & 12.1 & 76.2 \\
15 & 100 & 1.0 & 54.4 & 260.5 & 100 & 1.0 & 51.4 & 296.6 & 100 & 1.0 & 36.7 & 198.9 \\
20 & 100 & 1.0 & 88.7 & 784.2 & 98 & 0.99 & 94.1 & 983.5 & 100 & 1.0 & 73.5 & 634.7 \\
25 & 100 & 1.0 & 146.6 & 1812.8 & 91 & 0.94 & 158.6 & 2645.2 & 100 & 1.0 & 126.2 & 1227.1 \\
30 & 100 & 1.0 & 231.2 & 4769.1 & 82 & 0.88 & 212.9 & 6839.9 & 99 & 0.99 & 157.6 & 2531.5 \\
35 & 92 & 0.95 & 273.7 & 8543.7 & 69 & 0.77 & 293.7 & 9671.5 & 87 & 0.93 & 201.9 & 4672.9 \\
40 & 85 & 0.90 & 401.3 & 12391.4 & 53 & 0.62 & 421.2 & 16428.6 & 79 & 0.84 & 248.5 & 8740.1 \\
\hline
\end{tabular}}
\caption{Output statistics on directed diameter and average distance, and undirected irregularity.}
\label{tab:stats}
\end{center}
\end{table}

As we could have expected, system ``efficiency", \ie the average successes and execution times, widely depends on the studied problem. Moreover, the fact that irregularity is an invariant of lesser worst case complexity, and that this problem is stated on undirected graphs can explain the improved running time for this invariant.

Additionally, from this first table, we can see that \DG\ is scalable with graph order. Indeed, while with average distance the raw success decreases a lot past a threshold, the amortised success remains acceptable. This suggests that, when no optimal solution can be found, it is still possible to find an individual with a fitness close to the optimal solution.

However, we have to keep in mind that these statistics could be output since we know optimal values for each underlying problem. When it is not the case, \ie when \DG\ is actually used to study new problems or conjectures, it would be wise to restrain the system to graphs of smaller orders, as we have no optimality guarantee. Moreover, in every case, a generic crossover operator such as \textsf{Align-2Point} was enough to find optimal values. Literature shows however that sometime, \eg with graph colouring \cite{GenAlgColor}, design a custom crossover dedicated to the studied problem is necessary.

The following table show parameters variations effects, that is, how the genetic algorithm behaves when changing some of its core components, like initial population, crossover and selection operators. Concretely, Table \ref{tab:avg} illustrates this behaviour for average distance, computed for graphs of order 15. Each table entry denotes amortised success for the underlying parameters, computed over 100 tests. 

\begin{table}[!htpd]
\begin{center}
\scriptsize{
\begin{tabular}{|c|c||r|r|r|r|}
\hline
\multirow{2}{*}{Selection} & \multirow{2}{*}{Crossover} & \multicolumn{4}{c|}{Initial population} \\
\cline{3-6}
& & Random & Deg. Seq. & Size block & HoG \\
\hline\hline
\multirow{3}{*}{Direct Elim.} & 1-Point & 0.81 & 0.71 & 0.84 & 0.85\\
\cline{2-6}
& 2-Point & 0.84 & 0.78 & 0.90 & 0.90 \\
\cline{2-6}
& \textsf{Align-2Point} & 0.87 & 0.79 & 0.91 & 0.94 \\
\cline{2-6}
& \textsf{Align-Greedy} & 0.88 & 0.81 & 0.90 & 0.94 \\
\hline\hline
\multirow{3}{*}{Roulette Wheel} & 1-Point & 0.83 & 0.74 & 0.91 & 0.91\\
\cline{2-6}
& 2-Point & 0.89 & 0.81 & 0.93 & 0.95\\
\cline{2-6}
& \textsf{Align-2Point} & 0.92 & 0.88 & 0.98 & 1.0 \\
\cline{2-6}
& \textsf{Align-Greedy} & 0.95 & 0.87 & 0.97 & 0.98 \\
\hline\hline
\multirow{3}{*}{Stoch. Un. S.} & 1-Point & 0.81 & 0.76 & 0.91 & 0.91\\
\cline{2-6}
& 2-Point & 0.91 & 0.84 & 0.95 & 0.96 \\
\cline{2-6}
& \textsf{Align-2Point} & 0.95 & 0.89 & 1.0 &  1.0 \\
\cline{2-6}
& \textsf{Align-Greedy} & 0.95 & 0.86 & 1.0 & 1.0 \\
\hline\hline
\multirow{3}{*}{Tournament} & 1-Point & 0.85 & 0.73 & 0.88 & 0.89 \\
\cline{2-6}
& 2-Point & 0.88 & 0.82 & 0.91 & 0.93 \\
\cline{2-6}
& \textsf{Align-2Point} & 0.90 & 0.83 & 0.93 & 0.95 \\
\cline{2-6}
& \textsf{Align-Greedy} & 0.89 & 0.86 & 0.94 & 0.96 \\
\hline
\end{tabular}}
\caption{Algorithm behaviour on operator switching for diameter.}
\label{tab:avg}
\end{center}
\end{table}

We observe that a simply randomly generated population is usually less effective than a population generated in blocks of fixed size. This might mean that size matters more than isomorphism distribution. Moreover, we noticed with further experiments that forcing at least one graph of each possible size to be generated improves amortised success by about 4\%. These are the results detailed in Table \ref{tab:avg}. We also note that the \HOG\ improvement slightly increases the algorithm efficiency.

Another benefit of this improvement, for which statistics are not detailed here, is that it vastly decreases the average number of generations needed to find optimums the first time. More particularly, we observe improvements from about more than 4000 needed generations to less than 1500, considering the underlying problem.

On the other hand, generating graph from spreading over a coding in degree sequence is the least efficient initial population among the three other ones. So far, we assume that the naive filtering routine described in Section \ref{sec_ga} is inappropriate to generate a ``good" initial population. Future work trying to improve this random generation could increase the effectiveness of this generator.

Regarding crossover operators, we observe that 2-Point crossovers generally provide better results than 1-Point crossover. An important point however is that aligning graphs and then perform a 2-Point crossover over their encodings always improves the algorithm efficiency. However, for average distance, guiding the algorithm by choosing good subgraphs for the children does not seem to improve the algorithm. This could be explained by the fact that handling strong connectivity constraints in subgraphs penalises greatly optimisation, since these subgraphs will most likely not be strongly connected, especially smaller ones. As debated before, in problems such as graph colouring, this approach might however be extremely efficient.

Finally, roulette wheel selection seems to be more efficient than direct elimination. This could be explained by the fact that direct elimination is more (maybe too much) elitist in the way that, when using this routine, maximum is always selected while minimum is always dropped. Regarding the algorithm diversity and convergence, this can be a drawback in case of ``too deep" local optimum. Moreover, reducing bias in universal stochastic sampling slightly improves the algorithm efficiency. At last, we note that tournament selection usually provides better results than direct elimination, although it stays less efficient than roulette wheel selection. 

Similar tables for irregularity and diameter have not been provided since the system always find the optimal solution. More particularly, only switching initial population for these two invariants have a notable effect, and only on the time (in generations) needed to find the maximum the first time, in a range from 150 to 800 generations, not on the overall success.

%
%

\section{Concluding remarks and open questions} \label{sec_concl}

We considered a particular type of optimisation problem using graphs as \emph{feasible solutions} and not as instances, as it is often the case. They are most useful in extremal graph theory where they allow computer use in order to study problems and discover new results, such as conjectures, counterexamples, proof leads, etc. Numerous results output by AutoGraphiX~\cite{AGX1} justify this approach.

In this paper, we showed it was possible to design \emph{genetic algorithms} to solve these extremal graphs problems. Moreover, we described the system \DG\ implementing such algorithms and that is the first system to deal with directed graphs (as well as undirected ones). This system allows to find not only extremal graphs but also graphs satisfying given constraints, counterexamples and to check a graph transformation operator validity. Computational results show that this approach is working and relatively efficient. 

It is interesting to note that considered optimisation problems arise several fundamental questions and open problems regarding the use of genetic algorithms when individuals are (directed) graphs. These questions are not trivial since depend most of the time on the underlying studied problem, for instance on invariants defining the objective function or on graph constraints. We list here some of these questions :
\begin{itemize}
\item How to efficiently design objective function to deal with \emph{soft} constraints sensitive to crossover and mutation operators ?
\item How to define crossover and mutation operators preserving \emph{hard} constraints ?
\item How to define crossover operators handling heredity relevantly ?
\item How to design particular operators or encoding less sensitive to the studied problem~?
\item More generally, is it possible to design a genetic algorithm which is the most generic possible for any type of graph finding problem  ?
\item Would other metaheuristics be more suitable for this type of problems, other than VNS (AutoGraphiX) and genetic algorithms (\DG) ?
\end{itemize}

We gave some elements of answers to some of these questions although there are still many leads to follow, for instance regarding hybrids auto-adaptive algorithms which would automatically determine the best fitting operators (among available ones) for any given problem.

\section*{Acknowledgments}

Romain Absil has a Phd funded by the FRIA. We also thank Alain Hertz, for useful discussions about crossover design we use in this paper.


\begin{thebibliography}{10}

\bibitem{Albertson97}
{\sc Albertson, {\protect M.O}.}
\newblock {The Irregularity of a Graph}.
\newblock {\em Ars Combinatoria 46\/} (1997), 219 -- 225.

\bibitem{AGXsurvey}
{\sc Aouchiche, M., Caporossi, G., Hansen, P., and Laffay, M.}
\newblock {AutoGraphiX: a survey}.
\newblock {\em Electronic Notes in Discrete Mathematics 22\/} (2005), 515 --
  520.

\bibitem{Appel77}
{\sc Appel, K., and Haken, W.}
\newblock {Every Planar Map is Four Colorable. Part I. Discharging}.
\newblock {\em Illinois Journal of Mathematics 21\/} (1977), 429 -- 490.

\bibitem{Appel77b}
{\sc Appel, K., and Haken, W.}
\newblock {Every Planar Map is Four Colorable. Part II. Reducibility}.
\newblock {\em Illinois Journal of Mathematics 21\/} (1977), 491 -- 567.

\bibitem{Appel89}
{\sc Appel, K., and Haken, W.}
\newblock {Every Planar Map is Four Colorable.}
\newblock {\em Contemporary Mathematics 98\/} (1989), 1 -- 741.

\bibitem{Bang01}
{\sc Bang-Jensen, {\protect J.}., and Gutin, G.}, Eds.
\newblock {\em {Digraphs: Theory, Algorithms and Applications}}.
\newblock Springer, New York, 2001.

\bibitem{Brink03}
{\sc Brinkmann, G., Goedgebeur, J., M\'elot, H., and Coolsaet, K.}
\newblock House of graphs: a database of interesting graphs.
\newblock {\em Discrete Applied Mathematics 161\/} (2013), 311--314.

\bibitem{AGX1}
{\sc Caporossi, G., and Hansen, P.}
\newblock {Variable Neighborhood Search for Extremal Graphs 1. The AutoGraphiX
  System}.
\newblock {\em Discrete Math. 212\/} (2000), 29 -- 44.

\bibitem{Cvetkovic83d}
{\sc Cvetkovi\'c, D., and Kraus, L.}
\newblock {``Graph'' an Expert System for the Classification and Extension of
  the Knowledge in the Field of Graph Theory, User's Manual}.
\newblock Elektrotehn. Fak., Beograd, 1983.

\bibitem{Cvetkovic81}
{\sc Cvetkovi\'c, D., Kraus, L., and Simi\'c, S.}
\newblock {Discussing Graph Theory with a Computer, I: Implementation of Graph
  Theoretic Algorithms}.
\newblock {\em Univ. Beograd Publ. Elektrotehn. Fak, Ser. Mat. Fiz. No. 716 --
  No. 734\/} (1981), 100 -- 104.

\bibitem{Cvetkovic94}
{\sc Cvetkovi\'c, D., and Simi\'c, S.}
\newblock {Graph Theoretical Results Obtained by the Support of the Expert
  System ``Graph''}.
\newblock {\em Acad\'emie Serbe des Sciences et des Arts. Classe des Sciences
  Mathématiques et Naturelles 19\/} (1994), 19 -- 41.

\bibitem{Diestel00}
{\sc Diestel, R.}
\newblock {\em {Graph Theory}}, second edition~ed.
\newblock Springer-Verlag, 2000.

\bibitem{doyle02}
{\sc Doyle, J.~K., and Graver, J.~E.}
\newblock Mean distance in a directed graph.
\newblock {\em Environment and planning B 5\/} (1978), 19--25.

\bibitem{Fajtlowicz98}
{\sc Fajtlowicz, S.}
\newblock {Written on the Wall}.
\newblock A regulary updated file accessible from {\tt
  http://independencenumber.files.wordpress.com/2012/08/wow-july2004.pdf}.
  Available by request from S. Fajtlowicz at {\tt siemion@math.uh.edu}.

\bibitem{Fajtlowicz87}
{\sc Fajtlowicz, S.}
\newblock {On Conjectures of Graffiti -- II}.
\newblock {\em Congressus Numerantium 60\/} (1987), 187 -- 197.

\bibitem{Fajtlowicz88b}
{\sc Fajtlowicz, S.}
\newblock {On Conjectures of Graffiti -- III}.
\newblock {\em Congressus Numerantium 66\/} (1988), 23 -- 32.

\bibitem{Fajtlowicz90}
{\sc Fajtlowicz, S.}
\newblock {On Conjectures of Graffiti -- IV}.
\newblock {\em Congressus Numerantium 70\/} (1990), 231 -- 240.

\bibitem{Fajtlowicz95}
{\sc Fajtlowicz, S.}
\newblock {On Conjectures of Graffiti -- V}.
\newblock In {\em {Seventh International Quadrennial Conference on Graph
  Theory}\/} (1995), vol.~1, pp.~367 -- 376.

\bibitem{GenAlgColor}
{\sc Galinier, P., and Hao, J.-K.}
\newblock Hybrid evolutionary algorithms for graph coloring.
\newblock {\em Journal of Combinatorial Optimization 3}, 4 (1999), 379--397.

\bibitem{Garey79}
{\sc Garey, {\protect M.R.}., and Johnson, {\protect D.S.}.}
\newblock {\em {Computers and intractability. A guide to the theory of
  NP-completeness}}.
\newblock Freeman and Company, 1979.

\bibitem{Goldberg01}
{\sc Goldberg, D., and Miller, B.}
\newblock Genetic algorithms, tournament selection, and the effects of noise.
\newblock {\em Complex systems 8\/} (1995), 193--212.

\bibitem{Hakimi01}
{\sc Hakimi, S.}
\newblock On realizability of a set of integers as degrees of the vertices of a
  linear graph.
\newblock {\em Journal of the Society for Industrial and Applied Mathematics
  10\/} (1962), 496--506.

\bibitem{WhatForms}
{\sc Hansen, P., Aouchiche, M., Caporossi, G., M\'elot, H., and Stevanovi\'c,
  D.}
\newblock {What Forms Do Interesting Conjectures Have in Graph Theory?}
\newblock In {\em {Graphs and Discovery}}, {\protect Fajtlowicz, S. \emph{et
  al.}}, Ed., vol.~69 of {\em {DIMACS Series in Discrete Mathematics and
  Theoretical Computer Science}}. American Mathematical Society, Providence,
  2005, pp.~231 -- 252.

\bibitem{AvgDistForest}
{\sc Hansen, P., Hertz, A., Kilani, R., Marcotte, O., and Schindl, D.}
\newblock Average distance and maximum induced forest.
\newblock {\em Journal of Graph Theory 60\/} (2009), 31--54.

\bibitem{Hansen02}
{\sc Hansen, P., and M\'elot, H.}
\newblock {Computers and Discovery in Algebraic Graph Theory}.
\newblock {\em Linear Algebra and its Applications 356\/} (2002), 211 -- 230.

\bibitem{agx9}
{\sc Hansen, P., and M\'elot, H.}
\newblock {Variable Neighborhood Search for Extremal Graphs 9. Bounding the
  Irregularity of a Graph}.
\newblock In {\em {Graphs and Discovery}}, {\protect Fajtlowicz, S. \emph{et
  al.}}, Ed., vol.~69 of {\em {DIMACS Series in Discrete Mathematics and
  Theoretical Computer Science}}. American Mathematical Society, Providence,
  2005, pp.~253 -- 264.

\bibitem{Hansen01b}
{\sc Hansen, P., and Mladenovi\'c, N.}
\newblock {Variable Neighborhood Search~: Principles and Applications}.
\newblock {\em European Journal of Operational Research 130\/} (2001), 449 --
  467.

\bibitem{Havel01}
{\sc Havel, V.}
\newblock A remark on the existence of finite graphs.
\newblock {\em \v{C}asopis pro P\v{e}stov\'an\'i matematiky 80\/} (1955),
  477--480.

\bibitem{Mckay90}
{\sc McKay, {\protect B.D}.}
\newblock {Nauty User's Guide (version 1.5)}.
\newblock Tech. rep., Department of Computer Science, Australian National
  University, 1990.

\bibitem{Mckay98}
{\sc McKay, {\protect B.D}.}
\newblock {Isomorph-free Exhaustive Generation}.
\newblock {\em Journal of Algorithms 26\/} (1998), 306 -- 324.

\bibitem{GphDesc}
{\sc M\'elot, H.}
\newblock Facet defining inequalities among graph invariants: the system
  {GraPHedron}.
\newblock {\em Discrete Applied Mathematics 156\/} (2008), 1875 -- 1891.

\bibitem{Mladenovic97}
{\sc Mladenovi\'c, N., and Hansen, P.}
\newblock {Variable Neighbourhood Search}.
\newblock {\em Computers and Operations Research 24\/} (1997), 1097 -- 1100.

\bibitem{Grinvin}
{\sc Peeters, A., Coolsaet, K., Brinkmann, G., {\protect Van Cleemput}, N., and
  Fack, V.}
\newblock {GrInvIn in a nutshell}.
\newblock {\em Journal of Mathematical Chemistry 45\/} (2009), 471 -- 477.

\bibitem{Robertson97}
{\sc Robertson, N., Sanders, D., Seymour, P., and Thomas, R.}
\newblock {The Four-Color Theorem}.
\newblock {\em Journal of Combinatorial Theory, Series B 70\/} (1997), 2 -- 44.

\bibitem{newGraph}
{\sc {Stevanovi\'c, D. and Branko, V.}}
\newblock {An Invitation to newGRAPH}.
\newblock In {\em {Rendiconti del Seminario Matematico di Messina}}, vol.~9 of
  {\em 2}. 2003, pp.~211 -- 216.

\bibitem{Talbi01}
{\sc Talbi, E.-G.}, Ed.
\newblock {\em {Metaheuristics - From design to implementation}}.
\newblock Wiley, Hoboken, 2009.

\bibitem{Turan}
{\sc Tur{\'a}n, P.}
\newblock Eine {E}xtremalaufgabe aus der {G}raphentheorie.
\newblock {\em Matematikai \'es Fizikai Lapok 48\/} (1941), 436--452.

\end{thebibliography}
\end{document}